\documentclass[runningheads]{llncs}
\usepackage{graphicx}
\usepackage{hyperref}
\begin{document}
\title{Emerging AI Security Threats for Autonomous Cars – Case Studies}
%
%
\author{Shanthi Lekkala \inst{1}  \and 
Tanya Motwani \inst{1} \orcidID{0000-0002-0114-4538} \thanks{First and second author has contributed equally} \and 
Manojkumar Parmar  \inst{1} \orcidID{0000-0002-1183-4399} \thanks{Corresponding author} \and
Amit Phadke\inst{1}}
\authorrunning{S. Lekkala et al.}
%
\institute{Robert Bosch Engineering and Business Solutions Private Limited, Bengaluru 560095, India \\ 
\email{\{Shanthi.Lekkala,Tanya.Motwani, Manojkumar.Parmar, Amit.Phadke\}@bosch.com}\\
\url{https://www.bosch-india-software.com/}
}
\maketitle              
\begin{abstract}
Artificial Intelligence has made a significant contribution to autonomous vehicles, from object detection to path planning. However, AI models require a large amount of sensitive training data and are usually computationally intensive to build. The commercial value of such models motivates attackers to mount various attacks. Adversaries can launch model extraction attacks for monetization purposes or step-ping-stone towards other attacks like model evasion. In specific cases, it even results in destroying brand reputation, differentiation, and value proposition. In addition, IP laws and AI-related legalities are still evolving and are not uniform across countries. We discuss model extraction attacks in detail with two use-cases and a generic kill-chain that can compromise autonomous cars. It is essential to investigate strategies to manage and mitigate the risk of model theft.
\keywords{Artificial Intelligence \and AIoT Cycle \and Model Extraction \and Model Theft \and Adversarial Examples \and Security}
\end{abstract}

\section{Introduction}
\begin{figure}[t]
    \centering
    \includegraphics[width=\textwidth]{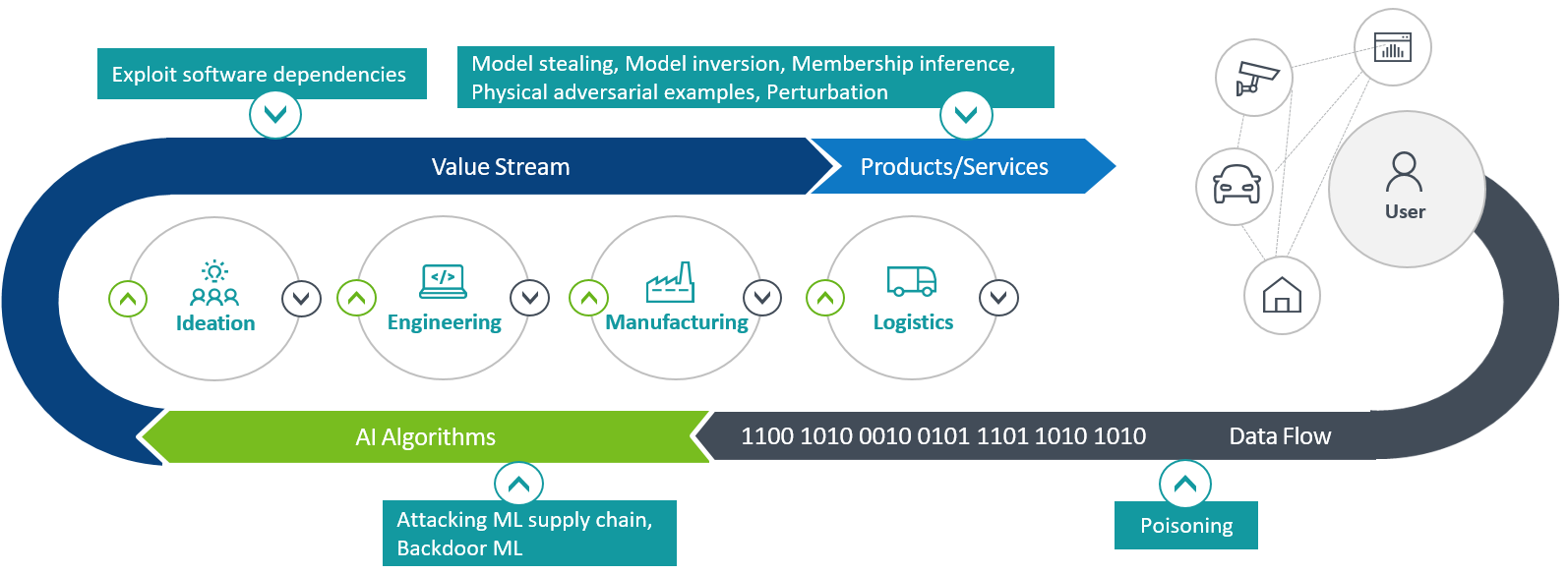}
    \caption{Emerging security threats across AIOT Cycle}
    \label{fig:fig1_aiot}
\end{figure}

Artificial intelligence has made significant breakthroughs and is widely used in various sectors such as self-driving cars, healthcare, agriculture, and others. There is a focus on numerous AI applications in the automotive world, from misfire detection for combustion engines, torque vectoring for electric vehicles to the camera, and LIDAR-based autonomous driving applications. Recent studies \footnote{There have been multiple attacks on ML systems of major companies such as Microsoft, Tesla, Google, and Amazon \cite{tencent_nodate,li_adversarial_2019,athalye_synthesizing_2018}. As per Gartner report \cite{gartner_ai_nodate}, by 2022, 30\% of cyber-attacks will involve data poisoning, model theft, or adversarial examples. This trend is only set to rise further.} showed that AI models are vulnerable to various attacks which can compromise products, services, and systems. Adversaries can steal high-value business models and offer them as a service at low cost to potential clients, leading to significant financial and intellectual property losses. They can also use model extraction to generate adversarial attacks and manipulate the outcome of the original model resulting in the ruination of brand reputation. Hence, it is vital to explore the risks and security threats involved in developing and deploying AI models.\\
The combination of AI and IoT (AIoT) allows us to build intelligent, connected, and autonomous systems. The AIoT cycle, as shown in Figure \ref{fig:fig1_aiot}, is a suitable representation to understand the data flywheel in autonomous systems. It demonstrates that more significant usage of systems (Products/Services) by users (User) results in more data (Data Flow). Accessing more data enables organizations to build better models (AI Algorithms) and ultimately a better autonomous system (Value Stream) to retain or get more customers. AIOT cycle promises faster and differentiated value creation. We discovered that the AIoT cycle is open to various security risks across its stages. Figure \ref{fig:fig1_aiot} illustrates 11 different threats \cite{kumar_failure_2019} across the AIOT cycle. The data flow phase is vulnerable to data poisoning attacks. AI algorithms are at risk to ML supply chain and backdoor ML attacks. Malicious users can exploit software dependencies in the value creation phase. In the products and services phase, systems are susceptible to perturbation, reprogramming of ML systems, recovery of training data, model extraction, physical adversarial attacks, model inversion, and member-ship inference attacks. This paper mainly focuses on model extraction attacks and their impact on autonomous cars related businesses
\section{Methodology}
\begin{figure}[t]
    \centering
    \includegraphics[width=\textwidth]{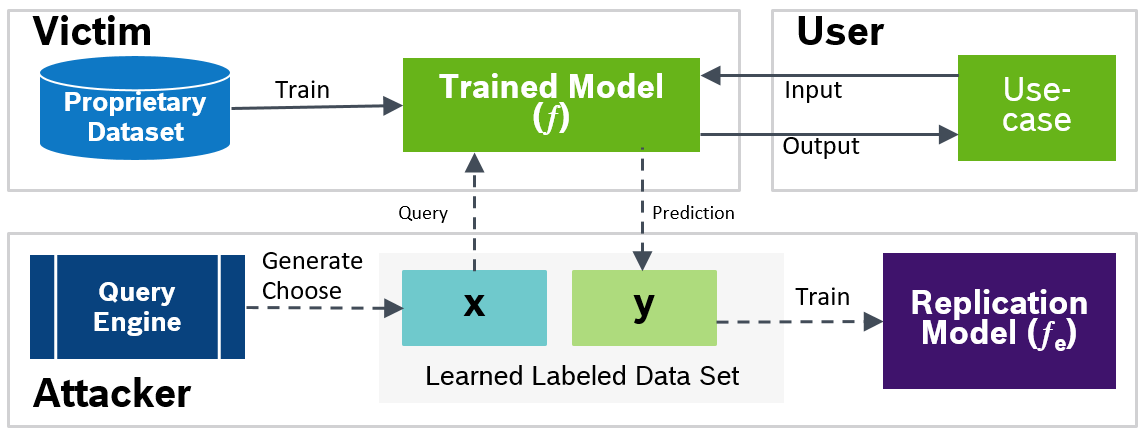}
    \caption{Intelligent Model Extraction Attack Framework}
    \label{fig:fig2_mea_framework}
\end{figure}

A model extraction attack (MEA) \cite{tramer_stealing_2016} is designed to duplicate the functionality (i.e., "steal") of the trained model and aims to reconstruct a local copy of the model. The concept of model extraction is depicted in Figure \ref{fig:fig2_mea_framework}. Consider an example where a company has trained an AI model using a proprietary dataset. An adversary can query($x$) this trained model ($f$) to obtain a prediction ($y$) on input feature vectors and, therefore, construct a learned labeled data set (\textit{LLDS}). The attacker can then train a replication model ($f_e$) on LLDS to approximate ($f$) without prior knowledge about its parameters. This mechanism can be viewed as a combination of two techniques: smart annotation and active learning transfer.\\
A model extraction attack is termed black-box if the adversary replicates the model using synthetic inputs without prior knowledge of domain or model parameters such as architecture, weights, and training dataset. This attack setting is closer to the real-world scenario.  It is gray-box if the adversary utilizes domain or model knowledge to generate synthetic inputs or chooses relevant samples from publicly available relevant datasets to query the model. As a result of the availability of additional information, gray-box attacks show better performance than black-box attacks to extract models\\
Authors have developed a technology \footnote{The software product AIShield based on patented technology described is being developed at our organisation and authors are primary members of the team. More details about product can be found at \href{https://www.bosch-india-software.com/en/products-and-services/innovation/cybersecurity-bosch-ai-shield/}{\textit{Link1}}, \href{https://azuremarketplace.microsoft.com/en-us/marketplace/apps/bosch.rbei_aishield?tab=Overview}{\textit{Link2}}}, to secure AI-powered products and services against model extraction attacks. It is built on a port-folio (patent-pending) of attack validation and defense generation mechanisms. We have identified more than 200 attacks based on AI models' input and output parameters to perform a vulnerability analysis using a proprietary attack framework. After vulnerability analysis of the model, AIShield generates an automated defense from 14 validated techniques and integrates it as a security layer for deployment in target systems.  
\section{Kill Chain – Autonomous Driving Vehicles}
\begin{figure}[t]
    \centering
    \includegraphics[width=\textwidth]{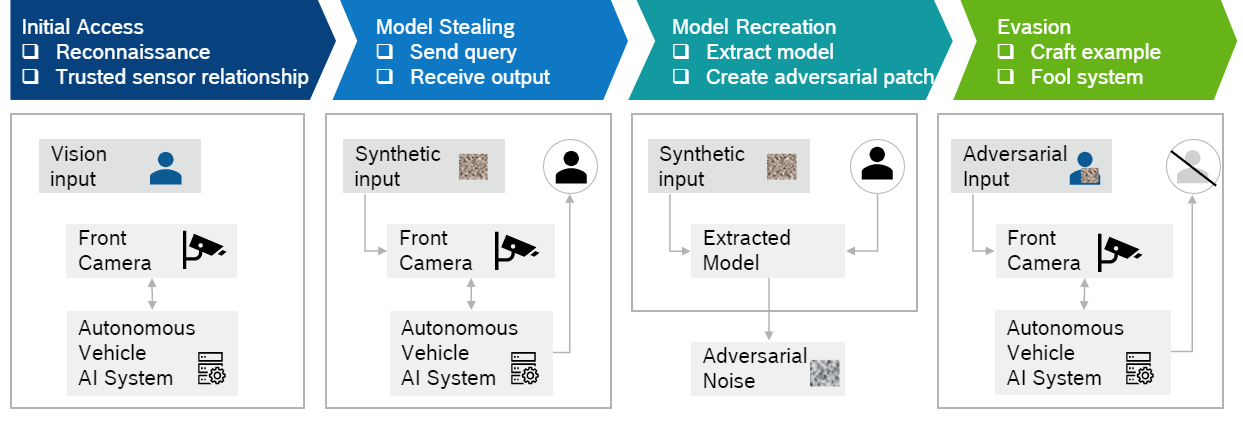}
    \caption{Killchain for Model extraction and Evasion}
    \label{fig:fig3_killchain}
\end{figure}

The focus of this study is to corroborate the role of model extraction in deceiving an AI model in the physical world. We intend to utilize model extraction to perform reconnaissance for adversarial attacks targeting the integrity of the model. \\
Consider the example of image recognition in an autonomous vehicle system (Figure \ref{fig:fig3_killchain}). Sensory data is captured by the front camera sensors, which AI models process for making decisions. We simulate this process through synthetic inputs generated by our intelligent attack engine to extract the model. The produced input-output pairs are used to train a functionally equivalent substitute model (i.e., the extracted model). The goal is to extract the model that achieves nearly 100\% agreement with the original model on the input space. In the next stage, the adversary uses gradients of the substitute model to generate adversarial examples. As we all know, it is challenging to conduct adversarial attacks on the model in a black-box setting. Through the process of model extraction, an adversary who initially operates in a black-box attack setting now works in a white-box attack setting. The adversary has recreated the original model parameters through the extracted model. The generated adversarial example significantly reduces the accuracy of the system. In such physical model extraction attacks, adversaries can utilize a large amount of data without any cost implications, resulting in highly accurate extracted models. The potential impact of this study is significant: self-driving vehicles can be vulnerable to such adversarial attacks and could fail to perform adequately.
\subsection{Pedestrian Detection (PD)}
\begin{figure}[t]
    \centering
    \includegraphics[width=\textwidth]{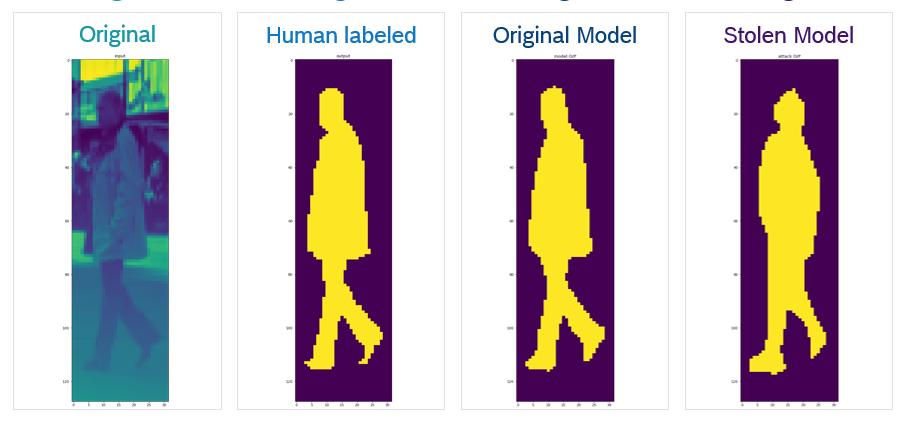}
    \caption{Results of Model Extraction for Pedestrian Detection Algorithm in Autonomous Driving Vehicles}
    \label{fig:fig4_padestrain}
\end{figure}

In an autonomous vehicle, one of the most important functionality is that of pedes-trian detection. We exemplify the kill chain using the case of pedestrian detection (as shown in Figure \ref{fig:fig4_padestrain}). We performed model extraction in a black-box attack setting using synthetic input-output pairs and achieved a stolen model IOU of 85.7\% with respect to the original model IOU of 96.1\%. We then conducted an evasion attack in a white-box setting using generated adversarial noise (FGSM \cite{goodfellow_explaining_2015}) on the extracted model. We were able to deceive the original model with 7\% of noise generated using the extracted model. Comparatively, 5\% adversarial noise (FGSM) was sufficient to fool the original system in a white-box attack setting with noise generated using the original model. This study confirms the transferability of adversarial examples in model extraction attacks. It also demonstrates that we can fool the system efficiently with only a minuscule difference in the percentage of attack noise. We used the gen-erated adversarial noise to launch data poisoning and membership inference attacks as well.
\subsection{Traffic Sign Recognition (TSR)}
In another relevant scenario, we experimented with the task of traffic sign recognition. Our setup involves an original model trained for traffic sign recognition on the GTSRB dataset \cite{stallkamp_man_2012}.  As the dataset consists of 43 classes, top-5 accuracy is used as a metric to measure the model's performance. The model follows a CNN architecture with a test accuracy of 97\%. We evaluate the effectiveness of both black-box and gray-box attack methods on the original model. In the black-box attack setting, queries are generated using random square blobs. Various noises are added to 3\% of images sampled from the original dataset in the gray-box attack setting. We build an attack dataset by combining the queries generated during the black box attack set-ting and these sampled images. This dataset is used to construct the input-output pairs that are used to extract the original model. The extracted model achieves a top-5 test accuracy of 91.69\% in the black-box attack setting and 94.47\% in the gray-box attack setting.
\section{conclusion}
With this study, we clearly outline the content of the planned presentation and its value to the ESCAR community. We present a comprehensive analysis of the model security threat in the products and services phase of the AIoT cycle. The expediency of model extraction attacks is demonstrated through case studies of pedestrian detection and traffic sign recognition. Model extraction is a credible threat to AI models that can have a significant financial, intellectual property, and brand impact for autonomous car OEMs and reduce trust towards the utility of autonomous cars in public perception. With the emergence of AIoT technology, it is imperative to focus on the security of autonomous systems to make them robust and safe for their adoption.
\bibliographystyle{splncs04}
\bibliography{references_ESCAR}
\end{document}